\documentclass[manuscript]{emulateapj}

\usepackage{rotating}

\newcommand{\bcen}{\begin{center}}
\newcommand{\ecen}{\end{center}}
\newcommand{\be}{\begin{equation}}
\newcommand{\ee}{\end{equation}}
\newcommand{\bdis}{\begin{displaymath}}
\newcommand{\edis}{\end{displaymath}}

\begin{document}

\title{Radiative transfer models of mid-infrared H$_2$O lines in the Planet-forming Region of Circumstellar Disks}

\author{R. Meijerink\altaffilmark{1},
  K.M. Pontoppidan\altaffilmark{1,2}, G.A. Blake\altaffilmark{1},
  D.R. Poelman\altaffilmark{3}, and C.P. Dullemond\altaffilmark{4}}

\altaffiltext{1}{California Institute of Technology, Division of
  Geological and Planetary Sciences, MS 150-21, Pasadena, CA
  9112}\email{rowin@gps.caltech.edu}

\altaffiltext{2}{Hubble Fellow}

\altaffiltext{3}{School of Physics and Astronomy, University of St
  Andrews, North Haugh, St Andrews KY16 9SS, Scotland}

\altaffiltext{4}{Max-Planck-Institut f\"ur Astronomie, K\"onigstuhl 17,
  69117 Heidelberg, Germany}

\begin{abstract}
  
  \noindent
  The study of warm molecular gas in the inner regions of
  protoplanetary disks is of key importance for the study of planet
  formation and especially for the transport of H$_2$O and organic
  molecules to the surfaces of rocky planets/satellites. Recent {\it
    Spitzer} observations have shown that the mid-infrared spectra of
  protoplanetary disks are covered in emission lines due to water and
  other molecules. Here, we present a non-LTE 2D radiative transfer
  model of water lines in the 10-36\,$\mu$m range that can be used to
  constrain the abundance structure of water vapor, given an observed
  spectrum, and show that an assumption of local thermodynamic
  equilibrium (LTE) does not accurately estimate the physical
  conditions of the water vapor emission zones, including temperatures
  and abundance structures. By applying the model to published {\it
    Spitzer} spectra we find that: 1) most water lines are
  subthermally excited, 2) the gas-to-dust ratio must be as much as
  one to two orders of magnitude higher than the canonical
  interstellar medium ratio of 100-200, and 3) the gas temperature
  must be significantly higher than the dust temperature, in agreement
  with detailed heating/cooling models, and 4) the water vapor
  abundance in the disk surface must be significantly truncated beyond
  $\sim 1~$AU. A low efficiency of water formation below $\sim 300\,$K
  may naturally result in a lower water abundance beyond a certain
  radius. However, we find that chemistry, although not necessarily
  ruled out, may not be sufficient to produce a sharp abundance drop
  of many orders of magnitude and speculate that the depletion may
  also be caused by vertical turbulent diffusion of water vapor from
  the superheated surface to regions below the snow line, where the
  water can freeze out and be transported to the midplane as part of
  the general dust settling. Such a vertical cold finger effect is
  likely to be efficient due to the lack of a replenishment mechanism
  of large, water-ice coated dust grains to the disk surface.
  
\end{abstract}


\section{Introduction}

Planets are thought to predominantly form within 10~AU of young low
mass stars ($<$ a few solar masses). Concurrent with the process of
planet formation, a rich carbon, nitrogen and oxygen chemistry evolves
to eventually form the building blocks of the icy bodies within young
planetary systems; moons around giant planets, comets and Kuiper belt objects. This
chemistry will also seed the surfaces of rocky terrestrial planets
with water and organics, probably either through impacts with comets
or carried in hydrated refractory meteorites \citep{Raymond2004}. 

This early, highly active and evolving stage takes place within the
first few million years of a star's lifetime while the protoplanetary
disk is still gas-rich. The planet forming region can therefore be
traced by observations of warm (100-2000~K) molecular gas, through a
multitude of rotational and rovibrational transitions spanning the
infrared wavelength range (2-200~$\mu$m).

One critical question is how the distribution of water occurred in the
Solar Nebula and how the surface of the young Earth was
hydrated. \citet{Salyk2008} and \citet{Carr2008} report detections of
hundreds of water vapor lines in three protoplanetary disks observed
with the {\it Spitzer} Space Telescope InfraRed Spectrometer (IRS)
between 10 and 36~$\mu$m. These first, pioneering spectra were
analyzed using simple `slab models' that assume level populations
consistent with thermodynamic equilibrium and otherwise consisting
only of a column density, a single excitation temperature and an
emitting solid angle.

Fits to data using such simple slab models result in column densities
of $N_{\rm CO}=6-7\times 10^{18}$~cm$^{-2}$, $N_{\rm H_2O}=6-8\times
10^{17}$~cm$^{-2}$, and $N_{\rm OH}=2\times 10^{17}$~cm$^{-2}$ and an
excitation temperature $T=1000$~K for DR Tau and AS 205A, which is
similar to the observations of AA Tau by \citet{Carr2008}. The latter
paper also reports the detection of CO$_2$ and that of the small
organics C$_2$H$_2$ and HCN. Given a slab model, the line emission is
derived to be confined within radii ranging from 0.6 to 3.0~AU.

Recently, a number of theoretical models have appeared that focus on
the gas-phase thermo-chemical modeling of the inner regions of
protoplanetary disks \citep{Markwick2002, Glassgold2004, Nomura2005,
  Nomura2007, Agundez2008, Gorti2008, Ercolano2008, Woods2009,
  Glassgold2009, Woitke2009}. These papers have different levels of
complexity in their adopted physics (e.g., treatment of X-rays and/or
FUV, the possibility of transport throughout the disk, time-dependent
treatment of chemistry) and deal with various chemical aspects (e.g.,
organic species, H$_2$O, carbon fractionation). Other theoretical
models focus on the dynamical transport of individual species through
the disk \citep{Stevenson1988,Ciesla2006,Ciesla2009}.  A key question
is therefore whether the large amounts of warm ($T\sim 300-1000$~K)
water that have been observed can be explained by models with either
{\it in situ} formation or {\it dynamical transport} from larger
radii.

To date, neither the observational work nor the theoretical papers
that consider the chemistry in the inner region of protoplanetary
disks have shown how physical parameters -- specifically the actual
{\it abundance structure} of a given molecule -- can be derived from
line observations in a meaningful way. In this paper, we propose that,
given the complexity of inner disk chemistry and volatile dynamics,
progress can best be made by separating the chemical-dynamical model
from the radiative transfer modeling of observations. This way, high
quality observations, such as infrared imaging spectroscopy, can be
used to {\it derive} an abundance structure of a given molecule by
application of a radiative transfer code that performs: 1) a non-LTE
excitation calculation in a disk-geometry followed by, 2) raytracing
that correctly samples the velocity field at small radii, and 3) a
front end that simulates the action of the appropriate
telescope/instrument. Given a derived abundance structure, the
interpretation step, using chemical-dynamical models, becomes much
easier than the approach in which one begins with the chemical
model. This paper describes step 1). A companion paper describes steps
2) and 3) \citep{Pontoppidan2009}.

Here our major focus is an examination of the excitation and infrared
radiative transfer of water vapor in the surface layers of
protoplanetary disk. We find that water vapor must be strongly
depleted from the surface over a significant range of radii. This can
be due to a chemical effect, as water is less efficiently formed at
temperatures $T<300$~K \citep{Glassgold2009}; but we speculate that
additional depletion is active due to vertical gas diffusion, followed
by freeze-out below the snow line and settling to the mid-plane on icy
dust grains. This process is analogous to the radial cold finger
effect proposed by \cite{Stevenson1988}, but will not be balanced by
the inward radial migration of icy bodies \citep{Salyk2008}. To
distinguish it from the radial cold finger effect, we call it the {\it
  vertical cold finger effect} (see Section
\ref{fiducial_model_description}).

To place the non-LTE model into context, we first describe slab models
of disk infrared emission and the collisional/radiative excitation of
water. The details of the statistical equilibrium calculations are
then presented, along with our initial results and general
conclusions. The models presented here are meant only to provide a
fundamental analysis of the global properties of water vapor emission
from disks, the more elaborate parameter study of the water excitation
model that is needed to fit the Spitzer IRS spectra of individual
sources is postponed to a separate paper (Meijerink et al. 2009, in
preparation).

\section{LTE slab models versus non-LTE 2D models}

Recent papers from \citet{Salyk2008} and \citet{Carr2008} have been
successful in obtaining good matches to H$_2$O line observations in
the Mid-Infrared (MIR) {\it Spitzer}-IRS band ($\lambda \sim 10-36
\mu$m) using single temperature, single column density LTE models with
no global kinematics, such as Keplerian motion. These fits have been
used to provide numerical estimates of the warm water column
densities. Since a key point of this work is to show why non-LTE
calculations are required to quantitatively interpret the observed
molecular line spectra of the warm inner regions of protoplanetary
disks, particularly the emission from H$_2$O, we first examine the
successes and limitations of LTE `slab' models.  We will argue below
that taking the properties derived from such models at face value is
dangerous and may well lead to erroneous conclusions when attempting
to compare abundances with chemical models. Specifically, a good fit
in a given wavelength range does not imply that the fitting model is
correct. The low resolution of the {\it Spitzer} spectra, and the
resulting lack of line profile information, introduce significant
degeneracies, as we discuss next. We will then show that the
introduction of relatively simple physics lead to constraints not
available from slab models, and that a requirement to match all lines
in the observed {\it Spitzer} range can significantly constrain the
results. Although the lineshapes also contain a great deal of info, it
is not considered here, since the IRS has a resolution of
$R=\lambda/\Delta\lambda=600$ (500 km/s).

Slab models ignore the complex geometric structure of protoplanetary
disks. In the inner regions, densities can be as high as $n_{\rm
  H}>10^{16}$~cm$^{-3}$ in the midplane and at small radii
($R<0.5$~AU), but decrease with radius and height (see
Fig. \ref{timescale_comp}) over many orders of magnitude to densities
$n<10^{3}$~cm$^{-3}$. Gas temperatures can be as high as $T\sim
5000~K$ in the upper reaches of the disk atmosphere where the X-ray
\citep{Glassgold2004} or FUV \citep{Kamp2004} attenuation is minimal.
The temperature drops steadily in the shielded regions of the disk to
values of $T\sim 10-100$~K, where the gas temperature couples strongly
to the dust temperature. Therefore, the disk consists of an ensemble
of densities and temperatures. Lines are, in general, the result of
emission from a range of disk radii, and thus cannot be represented by
a single excitation temperature. The mid-infrared wavelength region
contains radiative transitions from upper levels with an enormous
range of energies, from values at/near the photon energy ($E < 720$~K
or 500~cm$^{-1}$) to values in excess of $E>7200$~K or
5000~cm$^{-1}$. The conversion between wavenumber and temperature
units is 1.44. Temperature units are used throughout this paper.  For
a light, asymmetric top molecule, such as H$_2$O, transitions with
various upper state energies can be highly scattered throughout the
spectrum, and even a small spectral region can contain lines that
trace very different regions of the disk. Therefore, the assumption
that all lines are formed in the same region of the disk is
immediately seen to be inappropriate -- and can easily lead to an
erroneous determination of the line optical depth in a slab model. In
fact, it is essentially meaningless to even assign an optical depth to
a given line, since emission from different disk radii will have local
optical depths varying by orders of magnitude.

Furthermore, LTE assumes that the level populations are in accordance
with the Maxwell-Boltzmann distribution at a particular
temperature. This only holds when collisions dominate the molecular
excitation/relaxation, or numerically when the ambient densities are
higher than the critical densities $n_{crit}=A_{ul}/C_{ul}$, where
$A_{ul}$ and $C_{ul}$ are the radiative and collisional decay rates of
a particular transition. The critical densities for the excitation of
the water lines through collisions range between $n_{crit}=10^{8}$ to
$>10^{12}$~cm$^{-3}$, and tend to increase strongly with excitation
energy. Pure rotational transitions with low excitation energies,
which lie in the frequency range of the Heterodyne Instrument for the
Far-Infrared (HIFI) on the Herschel Space Observatory ({\it
  Herschel}), have critical densities in the range
$10^8-10^9$~cm$^{-2}$. Unfortunately, these lines will likely be
difficult to observe toward disks with HIFI for sensitivity reasons
\citep{Meijerink2008}.  {\it Spitzer}-IRS ($10-36$~$\mu$m range) data,
on the contrary, have shown rich spectra of warm water in the inner
regions of protoplanetary disks. These consist of rotational
transitions with even larger critical densities than the HIFI lines
and are therefore unlikely to be thermalized. Lower excitation lines
of a given species tend to have lower critical densities, and as a
result, lines with lower excitation energies will trace larger regions
of the disk.  It is therefore expected that the line shapes and line
images will differ significantly from one transition to another. Thus,
{\it future spectrally and spatially resolved data will be crucial for
  constraining the water abundance structure}, and should be
interpreted with tools that specifically model (or mimic) the physical
and chemical properties of the disk.

\section{Excitation of H$_2$O}

There are a number of ways to excite water to higher energy states: 1)
collisions with, e.g., H, H$_2$, He, and electrons; 2) excitation by
radiation produced by hot dust located at the inner rim of the disk;
3) photo-desorption from dust grains; and 4) chemical formation in
highly excited states. In this paper, we only take into account 1) and
2). For this study all hydrogen is assumed to be in molecular form,
and He excitation is considered by scaling the molecular hydrogen
rates. Electron excitation is neglected, since electrons are expected
to be abundant only in unshielded low density regions, where water is
subthermally excited and its fractional abundance very low, which
therefore do not contribute much to the overall line emission.

Water lines in the $10-36~\mu$m wavelength range covered by the {\it
  Spitzer}-IRS Short-High (SH) and Long-High (LH) modules are
overwhelmingly dominated by pure rotational transitions between levels
in the ground ($v_1v_2v_3$)=(000) vibrational state as well as between
those in the first vibrational state ($v_1v_2v_3$)=(010). Further, for
a non-LTE calculation, transitions that connect the two vibrational
levels must also be considered.

Given the widespread importance of water, there have been significant
efforts to calculate the H$_2$O excitation rates -- both
experimentally and theoretically. Unfortunately, most studies only
give state-to-state rate coefficients below the first excited
vibrational level, i.e, below $E=2294.4$~K \citep{Green1993,
  Phillips1996, Faure2004,Faure2007, Dubernet2002, Grosjean2003}. The
only H$_2$O-H$_2$ and H$_2$O-electron collisional rates for higher
levels of excitation currently available are from
\citet{Faure2008}. This database contains all collisional rotational
and ro-vibrational transitions between the 5 lower vibrational levels,
($v_1v_2v_3$)=(000), (010), (020), (100), and (001), up to an energy
$E=7200$~K. The authors expect the data to be accurate within a factor
of $\sim5$ for the largest rates ($\ge
10^{-11}$~cm$^3$~s$^{-1}$). Smaller rates, which also include the
rovibrational excitation/de-excitation, are expected to be accurate to
an order of magnitude. Summed rates and cooling rates have better
precision and the data are well suited for the modeling of emission
spectra.

\begin{figure}[!htp]
\centerline{\includegraphics[height=100mm,clip=]{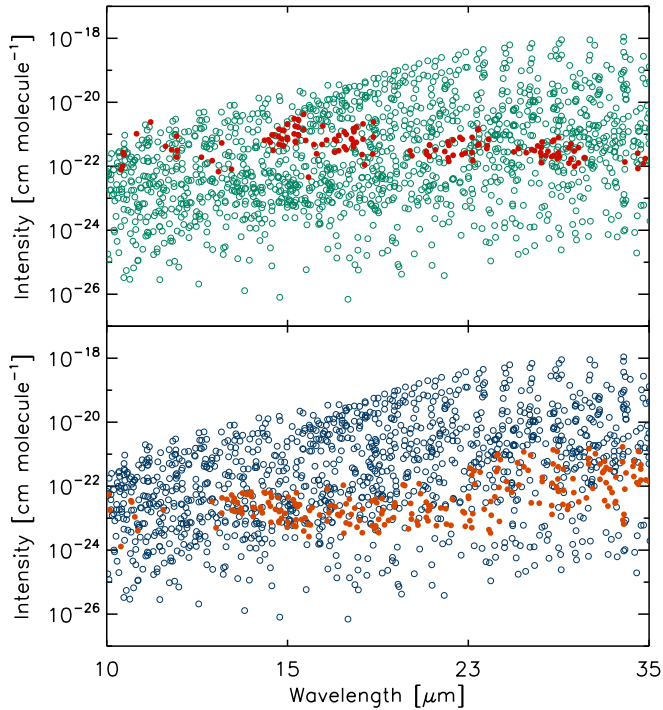}}
\caption{LTE intensities at $T=700$~K for: (top) Lines with upper
  level energy $E \le 7194$~K (open green circles) and upper level
  energy $E > 7194$~K (filled red circles); (bottom) Lines with lower
  and upper level in v=000 or v=010 (open blue circles) and all other
  transition (orange filled circles) both with upper level energy $E
  \le 7194$~K.}
\label{lines:HITRAN}
\end{figure}

In Fig. \ref{lines:HITRAN}, the intensities for lines at $T=700$~K are
shown, selected to be in the {\it Spitzer}-IRS spectral range between
10 and 36~$\mu$m. The top panel shows the intensity of all lines in
the HITRAN database \citep{Rothman2005}, where the lines with upper
level energy $E \le 7194$~K are the open green circles ($n=1454$), and
the lines with $E>7194$~K in red filled circles ($n=142$). Because
collisional rates between levels with energy $E>7194$~K are not
available it is possible that potentially strong lines are not
calculated, especially in the 10 - 15~$\mu$m spectral
range. Therefore, the calculation of these rates is important for
future modeling.

To limit the computational load, the non-LTE calculation is restricted
to transitions in the ($v_1v_2v_3$)=(000) and (010) vibrational
levels. In the lower panel in Fig. \ref{lines:HITRAN} we show the
intensity of transitions with upper level energy $E \le 7200$~K, where
the transitions between levels in the ($v_1v_2v_3$)=(000) and (010)
are in green open circles ($n=1230$) and all others ($n=224$) are in
red filled circles. It can be seen that the excluded vibrational
levels contain lines that are weaker by 1-2 orders of magnitude
relative to lines in the vibrational levels modeled. At the {\it
  Spitzer}-IRS SH+LH spectral resolution of R=600, these lines are
unlikely to contribute significantly to the overall spectrum.

\section{Model setup}

Next we investigate how the H$_2$O 10 - 36 $\mu$m spectrum depends on
the properties of the disk. The aim is to derive a fiducial model that
qualitatively matches the observed data. Quantitative fits to the
spectra of individual sources are postponed to a later paper. The
central star is chosen to be a representative T Tauri. It has mass
$M=1.0$~M$_\odot$, age $t=2$~Myrs, and solar metallicity $Z=0.02$,
corresponding to a radius $R=2.0$~R$_\odot$ and effective temperature
$T_{\rm eff}=4275$~K when using the pre-main sequence tracks of
\citet{Siess2000}. The stellar spectrum is a Kurucz stellar model
\citep{Kurucz1993}. An inclination of $30^\circ$ is adopted for
rendering of spectra.

The disk has a mass of $M=0.01$~M$_\odot$, outer radius $R_{\rm
  out}=120$~AU, flaring parameter $H/R\propto R^{2/21}$, and outer
pressure scale height $h_p / R = 0.10$. These parameters determine how
much light is intercepted at each radius and thus the temperature
distribution of the disk. We adopt a Gaussian density distribution in
the vertical direction and a radial surface density $\Sigma \propto
R^{-1}$.  The disk is roughly in hydrostatic equilibrium (for an
isothermal disk), but flatter than the \citet{Chiang1997} disk
($H/R\propto R^{2/7}$), giving us a better correspondence to the
observed spectral energy distributions in our sample of T Tauri disks.
However, we emphasize that the model is not an attempt to fit any
particular disk. The dust sublimation temperature is set at $T\sim
1600$~K, giving a radius of the inner rim of $R_{rim}=0.075$~AU.

In addition to parameters determining the dust structure of the disk,
a number of parameters are needed for the line modeling. These are the
prescription of the intrinsic line width, the freeze-out model, the
fractional H$_2$O abundance and the gas temperature and density
structure. We assume that the intrinsic linewidth throughout the grid
is determined by thermal broadening ($v_{th}=\sqrt{2kT / \mu_{\rm H2O}
  m_p}$), and turbulent broadening, assumed to be a fraction of the
sound speed $v_{turb}=\epsilon c_s$, where $\epsilon=0.03$ is adopted
in this paper. The intrinsic linewidth is therefore dominated by
thermal broadening. Five different model types are considered: 1)
Constant water abundance throughout the disk; 2) Reduced water
abundance in regions where water freezes out (see
Fig. \ref{lines:h2o_freeze_out}); 3) Decoupling of gas temperature and
destruction of water where gas is exposed to direct stellar
irradiation; 4) Increasing the gas-to-dust ratio from 1280 (which is
already and order of magnitude beyond the standard interstellar medium
value) to 12800, and 5) Truncation of the water abundance beyond
$R\sim 1$~AU due to the vertical cold finger effect. The model
parameters are summarized in Table \ref{model_params}.

\begin{table}[!h]
  \caption{Model input}
  \begin{center}
    \begin{tabular}{ll}    
      \hline
      \hline
      Parameter         & Computational range \\        
      \hline
      $v_{turb}=\epsilon c_s$ [km/s]        & $< 0.01 - 0.05$ \\ 
      $v_{th}=\sqrt\frac{2kT}{\mu_{H_2O}m_p} $ [km/s] & 0.1 - 2.0 (T=10-5000 K) \\
      $x({\rm H_2O})$                   & $10^{-10}-3\times 10^{-4}$  \\
      $T_{freeze}$                       & $100-200$~K (see Fig. \ref{lines:h2o_freeze_out})\\
      $H/R\propto R^{\alpha}$            & $\alpha=2/21$   \\
      $h_p/R$ (R=120~AU)                & 0.10 \\
      $M_*$ [M$_\odot$]                  & $1.0$  \\
      $R_*$ [R$_\odot$]                  & $2.0$ \\
      $T_{eff}$ [K]                      & $4275$ \\
      opacities                         & see Fig. \ref{grains:properties} \\
      gas-to-dust ratio                 & 1280 (1-3); 12800 (4-5) \\
      disk inclination                  & $30^\circ$ \\
      \hline
    \end{tabular}
  \end{center}
  \label{model_params}
\end{table}

\begin{figure*}[!htp]
\centerline{\includegraphics[height=75mm,clip=]{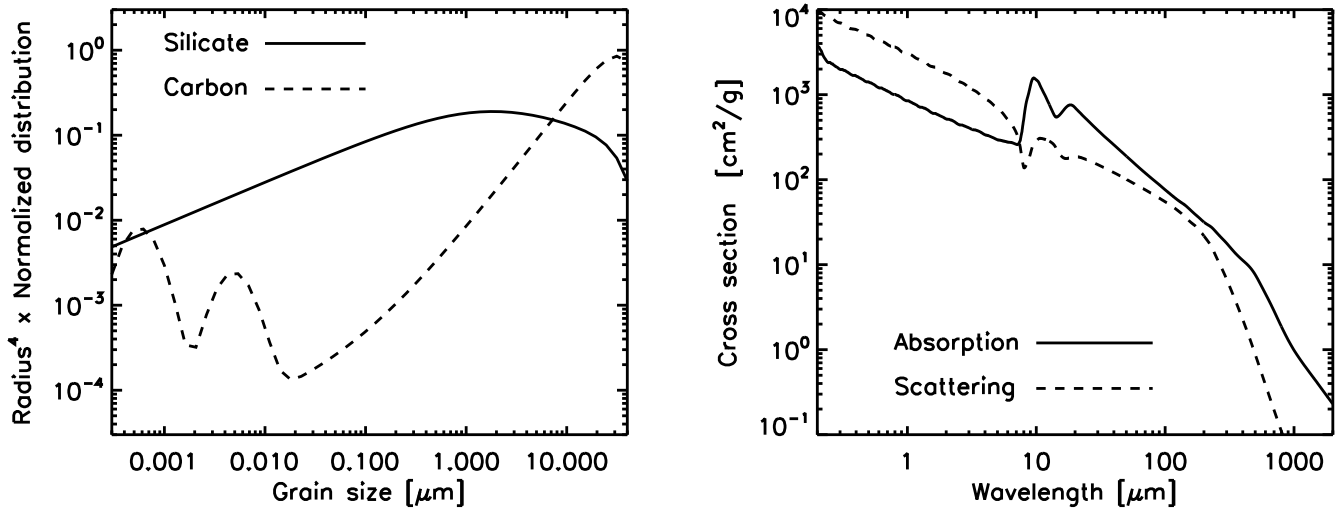}}
\caption{Dust size distribution with minimum grainsize
  $a_{min}=3\times 10^{-4}$~$\mu$m and maximum grainsize
  $a_{max}=40$~$\mu$m and opacities of the silicate (solid) and
  carbonaceous grains (dashed). Most of the silicate grain mass is in
  the larger grains ($R=0.5-20\mu$m), while the carbonaceous grains
  have substantial reservoirs in both in small ($R<0.01\mu$m) and
  large grains ($R>1\mu$m). }
\label{grains:properties}
\end{figure*}

The water emission spectrum is calculated as follows:

(i) First, the dust temperature distribution and the mean intensity in
the disk is determined using the 2D dust radiation transfer code RADMC
\citep{Dullemond2004}. The mean intensity is used to calculate the
radiative (de-)excitation rates $B_{ij}<J_{ij}>$.  The adopted dust
mixture contains 15 percent carbon grains and 85 percent silicate
grains. The two dust components are thermally coupled. The dust mass
distribution and opacities are shown in
Fig. \ref{grains:properties}. The silicate dust size distribution has
a peak at $1-10~\mu$m grains, while the carbon grains consists of both
small and very large grains ($>10~\mu$m). This corresponds to grains
that have grown beyond those found in the interstellar medium
\citep{Kessler2006}. It is implicitly assumed that grains larger than
40~$\mu$m have settled to the midplane where they do not affect the
water lines (which are formed near the disk surface).

(ii) The level populations are calculated with the radiative
excitation code $\beta$3D \citep{Poelman2005, Poelman2006}, a
multi-zone escape probability excitation code. It can be applied to
arbitrary geometries and is suitable for any atom and molecule.  This
code is adopted because it is about 10-100 times faster than existing
MC/ALI codes -- especially when the lines are highly optically thick
and IR pumped. While $\beta$3D can be used in 3 dimensional
structures, it is only feasible in applications where a limited amount
of levels ($\sim$10) are used. Here we adopt a 1+1D approximation to
alleviate the computational expense of iterating over a total of $\sim
540$ levels for both ortho and para water, especially for a large grid
of models. We calculate the excitation at a number of radial points
($\sim100-120$), each of which is effectively treated as a separate 1D
plane-parallel slab. The escape probability is therefore given by:

\begin{equation}
\beta(\tau)=\frac{1-\exp(-3\tau)}{3\tau},
\end{equation}

\noindent
where $\tau$ is the optical depth in the direction of the considered
ray. The resulting vertical excitation structures are subsequently
combined and regridded to polar coordinates, so that it can serve as
input for the next step: raytracing with RADLite.

(iii) RADLite (described in a companion paper, Pontoppidan et
al. \citeyear{Pontoppidan2009}) is used to obtain the line spectra. It
is a raytracer for axisymmetric geometries specifically set up to
handle the large velocity gradients and small turbulent widths which
are expected to be present in the planet-forming region of
circumstellar disks.

To demonstrate how the model can be used to investigate the influence of
different water abundance structures, two additional physical processes that 
determine the distribution of water vapor are included and described 
in the following sections. 

\subsection{Reduction of the water abundance due to freeze-out} \label{water_abundance}

Freeze-out of water onto grain surfaces can significantly reduce the
amount of water in the gas phase, and will reduce the integrated line
emission of the lower excitation lines \citep{Meijerink2008}. The
approach from \citet{Pontoppidan2006} is used to determine the
pressure and dust temperature where the transition from
water vapor to ice occurs on grains. The rate of ice mantle build-up
is given by:

\begin{equation}
\frac{dn_{ice}}{dt} = R_{ads} - R_{des},
\end{equation}

\noindent
where $R_{des}=\nu_0 \exp(-E/kT_{\rm dust}) \times n_{\rm H_2O, ice}
\times \gamma$ is the thermal desorption rate and $R_{ads}=n_{\rm
  H_2O} \times n_{dust} \times \pi d^2 \times \sqrt{3kT_{\rm
    gas}/m_{\rm H_2O}} \times f$. The surface binding energy of
$E=5773$~K is appropriate for water on a crystalline ice surface
\citep{Fraser2001}. The desorption and adsorption rates depend on the
dust and gas temperature, respectively, but freeze-out is only
important where the gas is coupled to dust, i.e., $T_{\rm gas}=T_{\rm
  dust}$. Non-thermal desorption mechanisms are not included. $\gamma$
is a factor that takes into account the observation (in the
laboratory) that H$_2$O only desorbs from the top monolayer (i.e.,
first order desorption for sub-monolayer coverage and zeroth order
desorption for multilayers), $d=1$~$\mu$m is the average adopted
grains radius, which is representative for that used by the radiative
transfer models, although the precise choice does not affect the
results, $f=1$ is the sticking coefficient, and $n_{\rm dust}$ and
$n_{\rm H_2O}$ are dust and water number densities. The
pre-exponential factor $\nu_0$ is expressed in s$^{-1}$ and is roughly
the frequency of the vibrational stretching mode ($\lambda
\sim$3.1~$\mu$m).

\begin{figure}[!htp]
\centerline{\includegraphics[height=70mm,clip=]{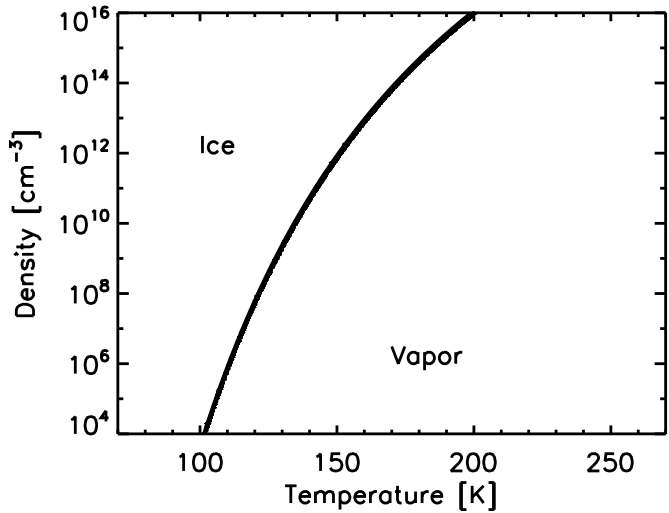}}
\caption{ The density dependence of the freeze-out temperature of
  H$_2$O.  Water is mostly frozen out for a give temperature and
  density left to the curve, and mostly in the gas phase when right to
  the curve}
\label{lines:h2o_freeze_out}
\end{figure}

In Fig. \ref{lines:h2o_freeze_out}, the density-dependent freeze-out
temperature is shown. The system is assumed to have reached
equilibrium, i.e., $dn_{ice}/dt=0$ -- a reasonable assumption because
the freeze-out timescales are short at the high densities in the disk
surface ($n \gtrsim 10^8$~cm$^{-3}$). At lower densities, $n=10^4 -
10^6$~cm$^{-3}$, where equilibrium is not reached on time scales
shorter than $10^{4-5}$ years and where photo-desorption occurs due to
direct irradiation from the central star, dust temperatures are
generally so high that freeze-out is not important. The freeze-out
temperature is between $T=100-110$~K for dense ISM cloud conditions,
$n=10^4-10^6$~cm$^{-3}$, and rises to temperatures as high as
$T=150-180$ K for midplane densities in the inner region,
$R=0.1-0.2$~AU, of the disk, $n=10^{12}-10^{14}$~cm$^{-3}$. If a
region is below the freeze-out temperature, we adopt a residual gas
phase abundance of $x_{\rm H_2O}=10^{-10}$ to simulate cosmic ray
desorption.

\begin{figure*}[!htp]
\centerline{\includegraphics[height=60mm,clip=]{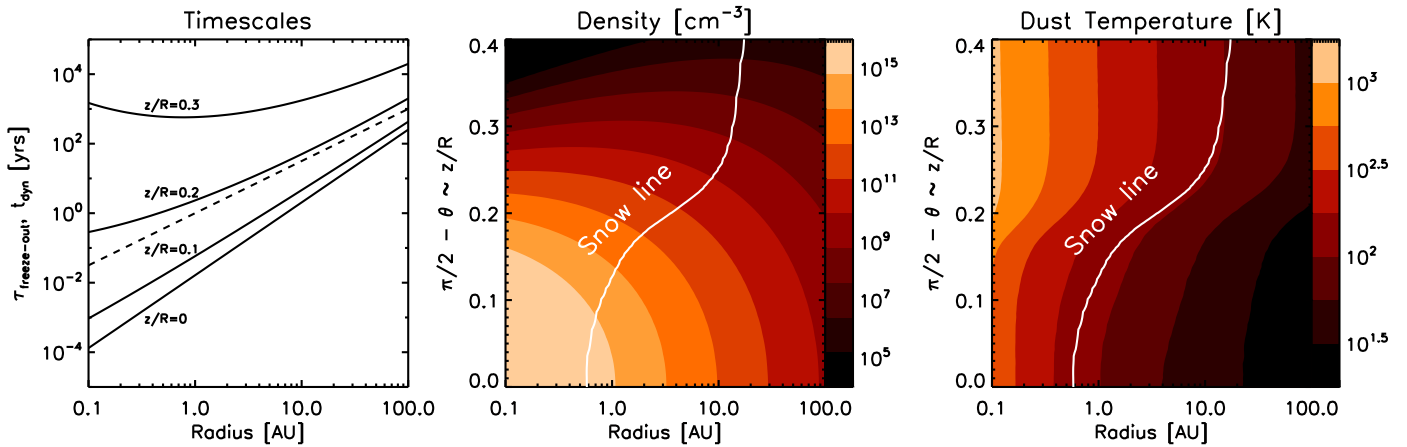}}
\caption{Left: The e-folding timescales for freeze out are indicated
  by the black curves for selected scale heights in the disk. The
  Keplerian dynamical timescale is indicated by the dashed line for
  comparison. Middle: Gas total hydrogen number density. The snow-line
  is indicated as well. Right: Dust temperature.}
\label{timescale_comp}
\end{figure*}

\subsection{Reduction of the H$_2$O abundance in the warm atmosphere} 
\label{water_destruction}

\begin{figure}[!htp]
\centerline{\includegraphics[height=60mm,clip=]{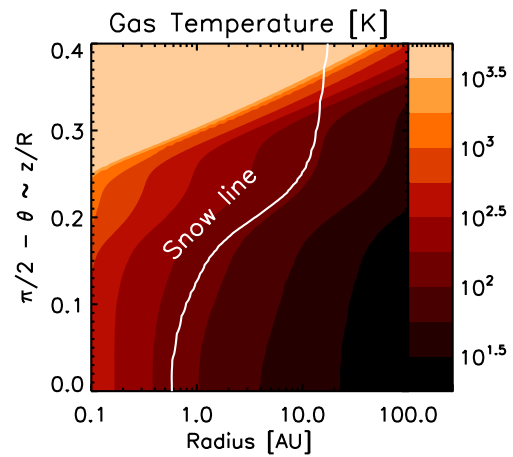}}
\caption{Gas temperature when heating of the upper atmosphere by
  X-rays is included (Model 5).}
\label{gas_temp}
\end{figure}

From static chemistry models \citep{Glassgold2004,Kamp2004,Woitke2009}
it is known that the gas and dust are kinetically uncoupled in the
warm atmosphere of the disk where the ambient densities are in the
range $n_{\rm H}=10^4-10^9$~cm$^{-3}$, leading to local gas
temperatures that are significantly higher than that of the dust.  The
gas is at high temperatures ($T\sim 5000$~K) in the uppermost
unshielded regions of the disk and drops to lower temperatures ($T\sim
100-200$~K) in deeper layers where the gas is coupled to the dust. If
this transition zone contains water, it will have a strong effect on
the line strengths of especially the high excitation lines. The
gas/dust decoupling is associated with a reduced water abundance in
the unshielded part of the disk due to photo-dissociation and
ion-molecule reactions. A simple parametrized description is adopted
here using the water abundance and temperature structure as shown in
Fig. 2 at radius $R\sim 1$~AU of \citet{Glassgold2009}. Warm water is
present at temperatures $T\sim 300-1000$~K, but reduced by orders of
magnitude at very high temperatures $T>2000$~K. The gas temperature
and water abundance depend solely on the ionization parameter
$\zeta/n_{\rm H}$, where $\zeta$ is the total ionization rate and
$n_{\rm H}$ the number density of hydrogen. The ionization parameter
is calculated assuming X-rays are the dominated radiation source. We
adopt a thermal source with $T_X=1$~keV and $L_X=10^{29}$~erg/s. The
total ionization rate for every point on the grid is calculated using
the absorption cross sections from \citet{Morrison1983}, and taking
into account the attenuating column $N_{\rm H}$. For every 1 keV of
energy absorbed, there are $1000 / 37 \sim 27$ ionizations (see
e.g. Glassgold et al. \citeyear{Glassgold1997} for more
details). Using this prescription, it is possible to scale the
\citet{Glassgold2009} results at $R=1$~AU to the entire disk given a
local ionization parameter, and the resulting gas temperature is shown
in Fig. \ref{gas_temp}. The adopted thermal and chemical profile can
of course be affected by vertical mixing, but such considerations are
beyond the scope of this paper.

\section{Results}

This section is divided into two different parts. The first part
highlights the differences in H$_2$O line spectra from the LTE and non-LTE
calculations for the same density and temperature
distribution, while Section \ref{fiducial_model_description} shows how the line 
spectra are affected by processes such as freeze-out, surface heating, 
chemistry and transport.

\subsection{LTE versus non-LTE}

In LTE the level population depends only on the temperature of the
gas. Conversely, in a non-LTE calculation all relevant collisional and
radiative excitation processes must be explicitly included. One
consequence of non-LTE is that levels may be subthermally excited when
the ambient densities are lower than the critical densities, or
superthermally excited when radiative excitation dominates the level
populations. When a level is subthermally populated in a particular
region of the disk, it has a smaller population than in LTE. The ratio
between the non-LTE and LTE level populations therefore indicates
whether a particular level is thermalized or not. A comparison of the
non-LTE and LTE face-on integrated column densities shows whether the
bulk of the gas at a certain radius is subthermally excited in a
particular level.

\begin{figure}
\centerline{\includegraphics[height=110mm,clip=]{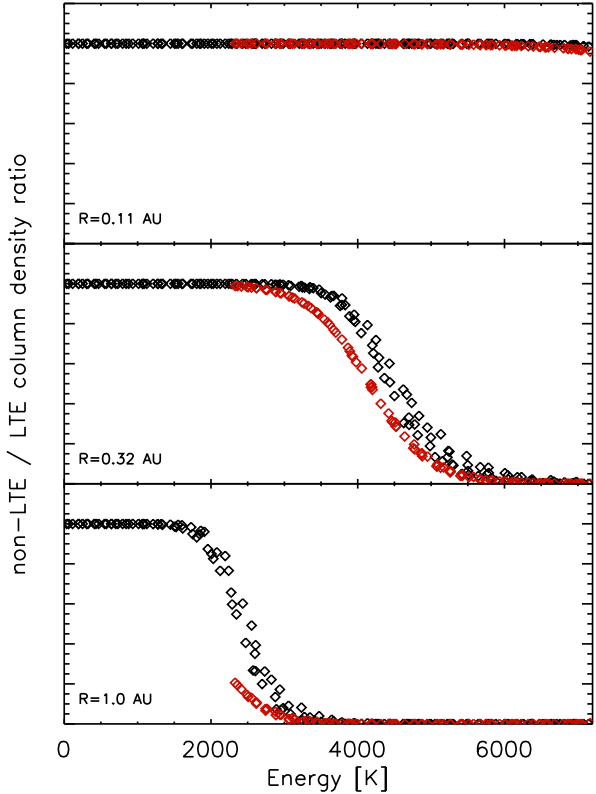}} 
\caption{The non-LTE / LTE integrated column density ratios for radii
  $R=0.11, 0.32$, and $1.0$~AU, for levels in the ground (black) and
  first (red) vibrational states.}
\label{weighted_columns}
\end{figure}

Fig. \ref{weighted_columns} shows the non-LTE / LTE integrated column
density ratios for all ortho-H$_2$O levels with energies $E<7194$~K at
three different radii with the same disk temperature and density
distribution and a constant water abundance $x_{\rm
  H_2O}=3\times10^{-4}$. It is immediately clear from this figure that
the LTE approximation is only valid at very small radii
($R\lesssim0.2$~AU). At larger radii the levels become sub-thermally
populated, especially the high excitation $E>2000$~K lines. The
subthermal decrease in column density is larger for levels with a
higher excitation energy. Because both density and temperature
decrease with radius, this results in a smaller emitting radius for
lines emitting from high excitation levels. The collisional excitation
rates $C_{ul}(T)$ decrease strongly with temperatures from $T=2000$ to
$200$~K in the disk, in some cases by more than two orders of
magnitude. This leads to a steep increase of the critical densities
with radius. Radiative excitation in part counteracts the lower
collisional excitation rates in the regions where the gas becomes
subthermally populated, but for the models in this paper, it does not
result in a significant increase in the integrated line flux.

\begin{figure}
\centerline{\includegraphics[height=90mm,clip=]{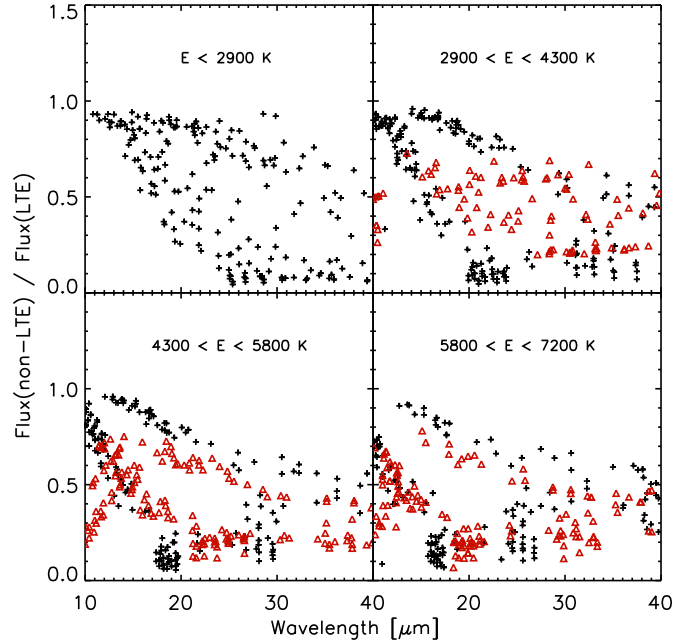}} 
\caption{Ratios of the non-LTE and LTE velocity integrated line fluxes
  for line with upper energy $E<2900$~K (top left), $2900<E<4300$~K
  (top right), $4300<E<5800$~K (bottom left), and $E>5800$~K (bottom
  right) in the {\it Spitzer}-IRS band. A distinction is made between
  the ground vibrational level (black crosses) and first vibrational
  level (red triangles).}
\label{line_ratios}
\end{figure}

Fig. \ref{line_ratios} shows the ratios between the non-LTE and LTE
velocity integrated line fluxes in the {\it Spitzer}-IRS band. The
points are marked with four different colors depending on their upper
level energy. The ratios are generally less than unity, meaning that
an assumption of LTE will overestimate line fluxes. The figure
illustrates that: 1) the spread in the ratios is very large, 2) the
lower the upper level energy $E_{up}$, the closer the ratio is to
unity, 3) the overall shapes of the LTE and non-LTE line spectra (when
viewed as an ensemble over the entire mid-infrared range) are very
different (see also Fig. \ref{panel_spectra}), and (4) the ratio shows
only a shallow trend with wavelength. The last point is an indication
that the upper level energies are evenly distributed with line
frequency thanks to the asymmetric top nature of the water
spectrum. This is particularly convenient from an observer's
perspective as it will usually be possible to observe lines with a
wide range of properties even if only a limited wavelength range is
available.

\begin{figure}
\centerline{\includegraphics[height=90mm,clip=]{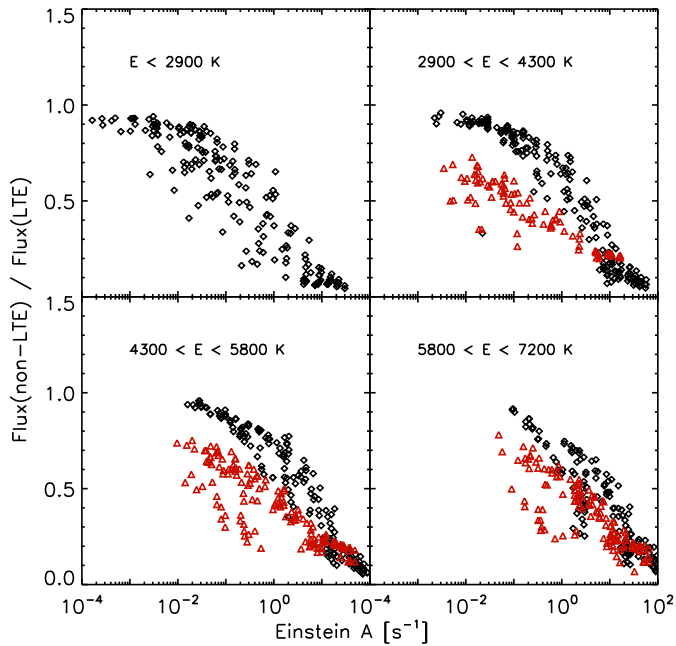}} 
\caption{Einstein A coefficients versus ratios of the non-LTE and LTE
  velocity integrated line fluxes for line with upper energy
  $E<2900$~K (top left), $2900<E<4300$~K (top right), $4300<E<5800$~K
  (bottom left), and $E>5800$~K (bottom right) in the {\it Spitzer}
  IRS band. A distinction is made between the ground vibrational level
  (black diamonds) and first vibrational level (red triangles).}
\label{line_ratios_EinsA}
\end{figure}

Fig. \ref{line_ratios_EinsA} shows the Einstein A coefficients versus 
the non-LTE/LTE line flux ratios, again separated into 
four groups as in Fig. \ref{line_ratios}. The figure illustrates the
following:
\begin{itemize}
\item[(i)] The line flux ratios decrease for larger Einstein A
  coefficients. This occurs because the critical densities are higher
  for transitions with larger Einstein A coefficients (the
  de-excitation rates are comparable when the upper level energies are
  similar).  Such transitions are therefore farthest from LTE.
\item[(ii)] For this particular model, lines with Einstein A
  coefficients below $\sim 10^{-2}$~s$^{-1}$ are close to LTE.
\item[(iii)] The ratios for transitions within the ground are higher
  than the first vibrational level. This is an indication that
  collisions dominate the excitation. If fluorescent excitation
  through the first vibrational bending mode were dominant then the
  line fluxes (at the same energy) become significantly larger for
  transitions from the first vibrational level that for those from the
  ground state.
\end{itemize}

\subsection{Toward a fiducial model} \label{fiducial_model_description}
\begin{sidewaysfigure*}
\centerline{\includegraphics[height=140mm,clip=]{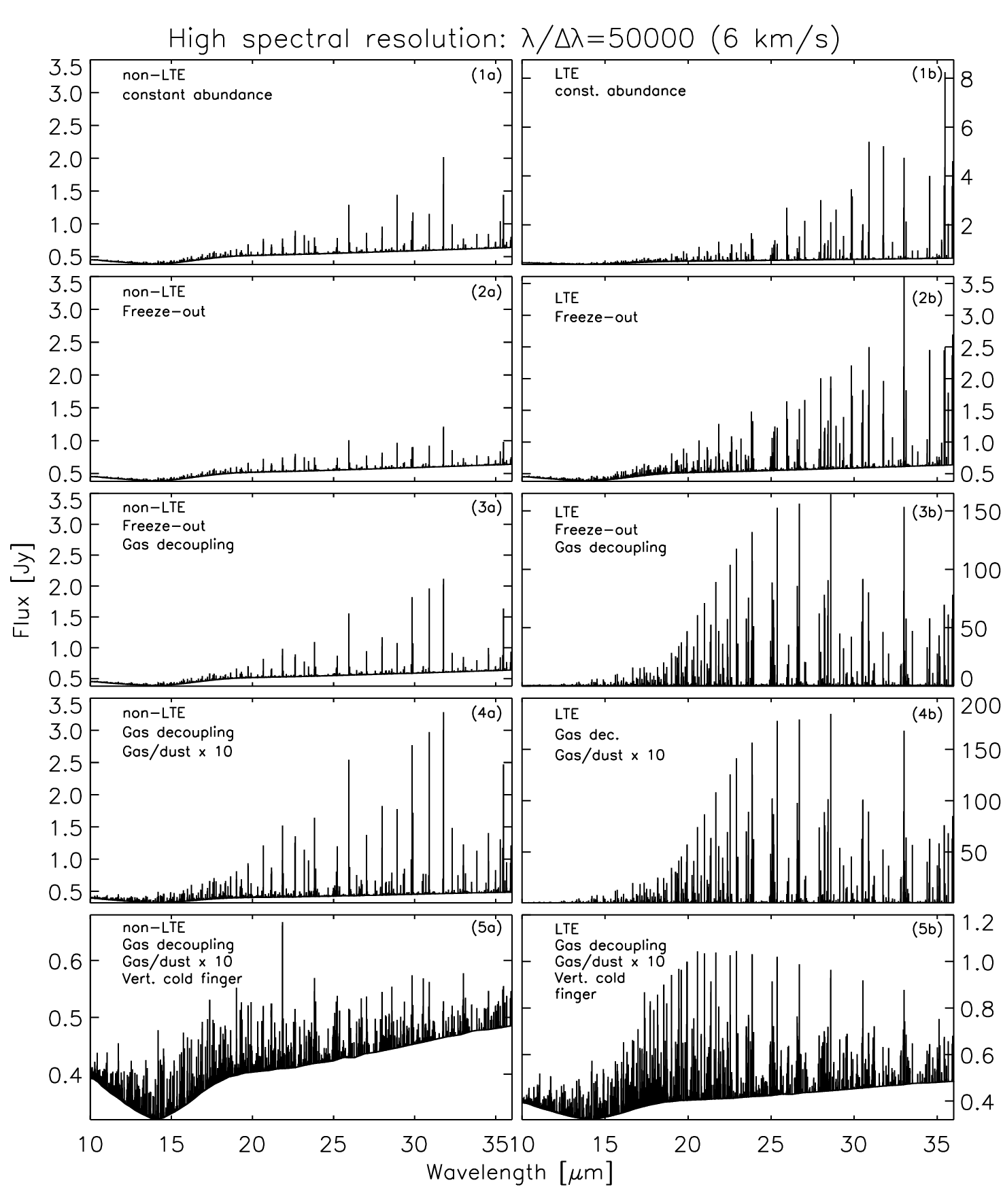}
\includegraphics[height=140mm,clip=]{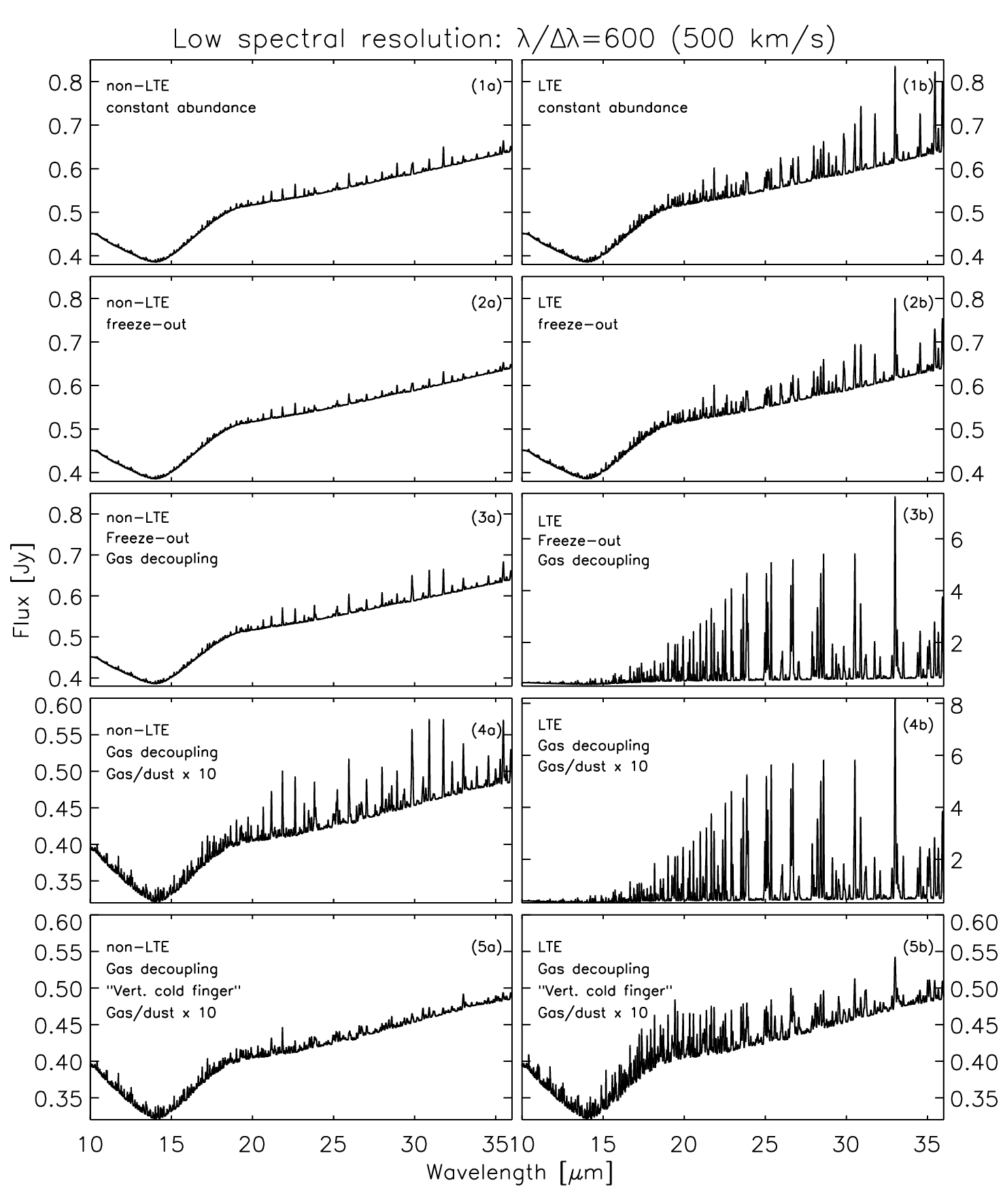}} 
\caption{Comparison of non-LTE (column a) and LTE (column row) line
  spectra for constant abundance distribution (model 1), then adding
  density dependent freeze-out (model 2), decoupling gas from dust in
  warm atmosphere (model 3), increasing the gas to dust ratio from
  1280 to 12800 (model 4), and reducing the H$_2$O abundance above
  mid-plane snowline, proposed as vertical cold finger effect (model
  5), respectively.}
\label{panel_spectra}
\end{sidewaysfigure*}

In this section a model that qualitatively reproduces the
shape of the observed H$_2$O line spectra is constructed.
Fig. \ref{panel_spectra} shows the line spectra for five different
models of non-LTE and LTE spectra convolved to two different
resolutions, $\lambda / \Delta \lambda =600$ (500 km/s, {\it Spitzer}
IRS) and 50000 (6 km/s).  All models have the same gas density
distribution (see Fig. \ref{timescale_comp}), but have an increasing
level of complexity of the water distribution and excitation to
accommodate the observables.

Model 1 is the reference model with a gas temperature equal to the
dust temperature and a constant water abundance ($x_{\rm H_2O}=3\times
10^{-4}$) throughout the disk, even in regions where water should
freeze out. In Model 2 a density dependent freeze-out of water has
been added (see Fig. \ref{lines:h2o_freeze_out}, and Section
\ref{water_abundance}). The main difference in the line spectra
between Model 1 and 2 is that some of the low excitation lines
decrease in intensity. The overall line spectrum does not change
drastically in the {\it Spitzer} range with freeze-out added
(especially in the non-LTE case). This is due to the relatively high
excitation temperatures of the lines from $10-36$~$\mu$m as well as
the fact that these transitions are subthermally excited at the
location of the surface snow line.  However, the freeze-out does
affect lines of lower excitation, located in the spectral regime
between 60 and 200$\mu$m \citep{Meijerink2008}, relevant for
Herschel-PACS. Although Model 1 and 2 are able to produce bright lines
of low excitation energy, similarly bright high excitation lines are
also required by the observed spectra, inconsistent with these models.

One way of increasing the strength of the high excitation lines is to
consider the decoupling of the gas temperature from the dust in the
upper part of the atmosphere of the disk. The parametrized treatment
of this decoupling, as discussed in Section \ref{water_destruction},
is added to Model 2, producing Model 3. Both the high and low
excitation lines increase in intensity (between 20-100 percent), but
the line to continuum ratios do not approach the observations, and the
low excitation lines are in fact boosted more than the high excitation
lines. Furthermore, the line-to-continuum ratio is still a factor of
3-5 too low with respect to representative {\it Spitzer} IRS
spectra. For instance, the prominent 33\,$\mu$m line complex now has a
line to continuum ratio of 0.05:1 at {\it Spitzer}-IRS resolution,
compared to observed values of $\sim$0.1-0.2:1.

In order to increase the general line-to-continuum ratio, Model 4 is
introduced, which is the same as Model 3, except that the gas-to-dust
ratio has been increased by factor of 10 to simulate strong dust
settling. The result is that higher gas densities contribute near the
disk surface leading to increased line intensities, coupled with a
decrease in the dust continuum. This model shows a much better
agreement with observations, but many low excitation lines are now
too bright. Furthermore, the line spectrum shows a strong increase in
peak line strength with wavelength, in contradiction with the
observed, almost flat, line integrated intensities.

The reason that the low excitation lines are bright is that they are
produced over a wide range of radii in the disk. Hence, they can be
suppressed if the water vapor abundance is drastically lowered beyond
a certain radius. {\it Static} chemical models do predict that the
water abundance is lowered by 1-2 orders of magnitude at temperatures
below $\sim 300~$K, which might be enough to reproduce some of the
observed line spectra, but due to the high optical depths that remain
at low temperature, the data indicate that higher depletions
may be necessary. Thus, other mechanisms should be
investigated. \cite{Stevenson1988} suggested that water vapor (and
therefore oxygen) is depleted from the inner disk, within the midplane
snow line, on time scales shorter than the life time of the disk. In
this mechanism the water vapor is transported from just within the
snow line (closer to the central star) to just outside the snow line
by turbulent diffusion, at which point the water freezes
out. Essentially, the sharp change in the partial pressure of water
vapor across the snow line results in a net flux of water molecules
outwards in the disk. However, this radial ``cold finger'' effect is
likely balanced by the inward migration of icy bodies, as discussed in
\citet{Salyk2008} and \citet{Ciesla2009}.

\begin{figure}[!ht]
\centerline{\includegraphics[height=70mm,clip=]{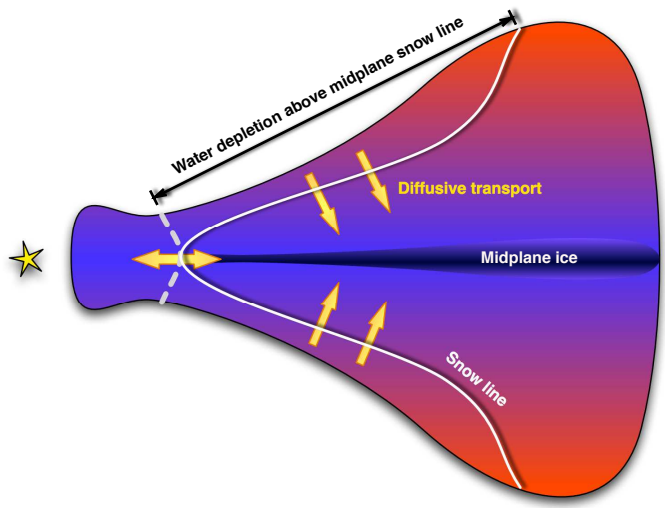}} 
\caption{Vertical cold finger effect: Water vapour will diffuse to
  regions below the snowline, where it will freeze out and take part
  in the general settling of dust to the midplane. For the fiducial
  model, the inner rim and the mid-plane snowline are located at
  R=0.07 and 0.7~AU, respectively.}
\label{vert_coldfinger}
\end{figure}

The radial cold finger effect assumes that the snow line is a vertical
line in an isothermal disk. However, due to the temperature inversion
in the disk atmosphere \citep{Glassgold2004, Kamp2004, Woitke2009},
the snow line is actually a curve that becomes almost parallel to the
disk surface over a very wide range of the disk, typically the entire
planet-forming region (0.5-20~AU). This can be seen in
Fig. \ref{timescale_comp}, and is sketched in
Fig. \ref{vert_coldfinger}. Hence, near the surface, the cold finger
effect will operate in the vertical direction; water vapor will
diffuse via turbulence to regions below the snow line where it will
freeze out and take part in the general settling of dust to the
midplane in a process one might call the ``vertical cold finger''
effect. The important difference relative to the {\it radial} cold
finger effect is that in the surface water vapor abundance will not be
replenished by the migration (in the case, the relofting) of solids if
icy dust grains grow to significant size as they settle and as seems
to be required by the significant millimeter-wave fluxes of many
disks, particularly those that are referred to as transitional disks
\citep[see][]{Brown2007b}. Further, the turbulence will be higher in
the surface, thus shortening the water depletion time
scale. Therefore, we speculate that such a mechanism is an effective
means of depleting surface water {\it above the midplane snow
  line}. If true, this would be of considerable importance because it
would mean that the radial distribution of surface water, as observed
with {\it Spitzer}-IRS and future mid- to far-infrared facilities, may
well trace the midplane snow line. More detailed modeling is needed to
estimate whether the time scales for this proposed process are short
enough to operate efficiently. It can be noted that a vertical cold
finger effect truncation above the snow line at $\sim 1~$AU neatly
explains why the slab models fit so well when assuming the same
emitting area for all water lines over a wide range of excitation
energies.  In this way, the water emission radius derived from slab
models may in fact be a measure of the location of the midplane snow
line.
   
\begin{figure*}[!ht]
\centerline{\includegraphics[height=120mm,clip=]{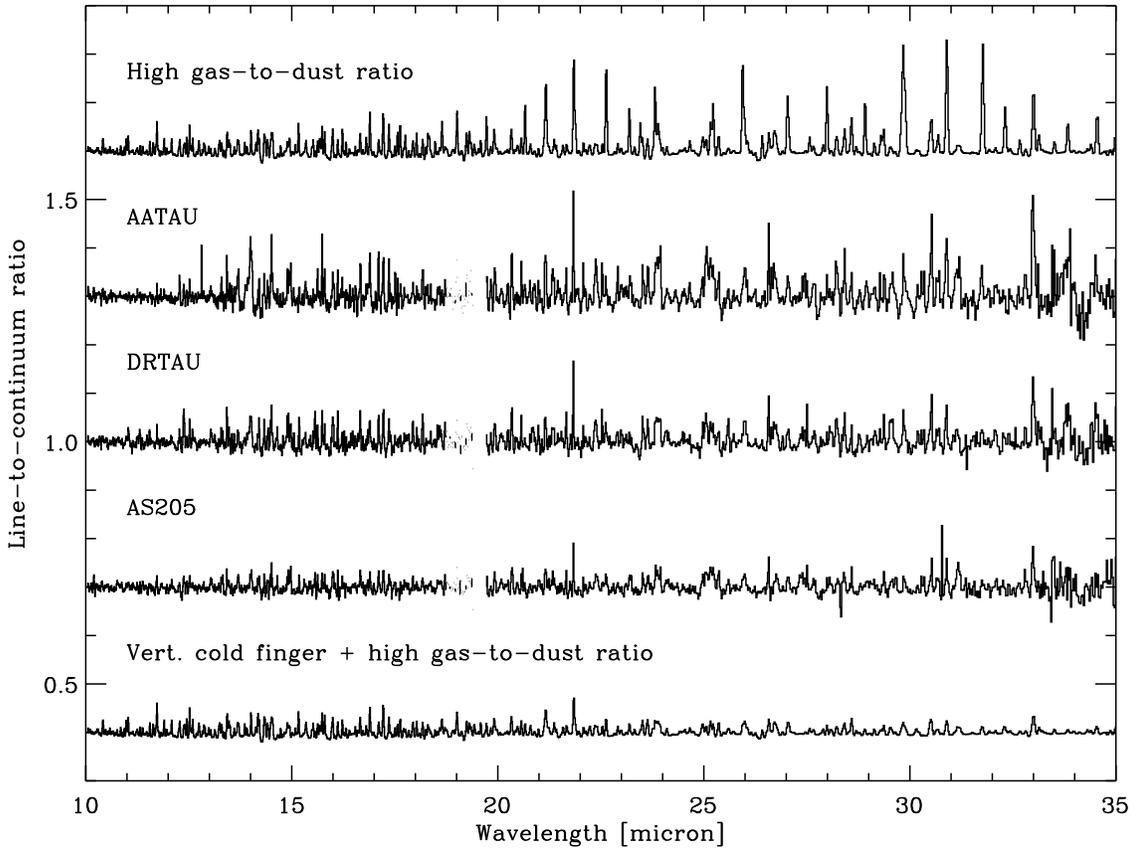}} 
\caption{Comparison of the 10-36\,$\mu$m model spectra to that of AA
  Tau, DR Tau and AS 205A, at R=600 \citep{Carr2008,Salyk2008}. The
  spectra are normalized to the continuum.}
\label{bracket}
\end{figure*}

In Model 5 we mimic the vertical cold finger effect by setting the
water abundance to the freeze-out abundance ($10^{-10}$) above the
midplane snow line of the disk, as illustrated in
Fig. \ref{vert_coldfinger}.  The resulting spectrum shows that the
line spectrum has significantly flattened, in general agreement with
the observed water spectra (see also Fig. \ref{bracket}).

\section{Discussion \& Conclusions}

Driven by the recent detection of hundreds of water emission lines in
the mid-infrared from disks around T Tauri stars by \citet{Carr2008}
and \citet{Salyk2008}, a non-LTE 2D model has been constructed that
qualitatively reproduces the observed spectra. The model can be
adjusted to obtain detailed fits to individual mid-infrared
spectra. Our approach is to derive an abundance structure for water
based on observational constraints, rather than use the prediction of
existing chemistry or transport models. As a demonstration of this
technique, the action of a vertical cold finger effect is proposed to
model the observed truncation of the surface abundance of water vapor.

The four most important conclusions are:

1) A non-thermal excitation treatment of water is essential in
determining the water distribution given an observed mid-infrared
water spectrum.  Assuming LTE in the calculation of H$_2$O line
spectra is generally invalid, especially for high excitation lines.
This is best illustrated in Fig. \ref{weighted_columns} and
\ref{panel_spectra}. The critical densities for exciting the water
lines in the 10 to 36\,$\mu$m region range from
$n_{crit}=10^8-10^{12}$~cm$^{-2}$, and rapidly increase with the upper
level energy. In Models 1 and 2, the difference between the LTE and
non-LTE dominant line intensities is a factor of 2 to 3.  In these
models, the gas temperature is assumed to be equal to that of the
dust, leading to faint high excitation lines. In Model 3, the surface
gas temperature is increased using a parametrization based on detailed
models of the gas heating \citep{Glassgold2009}. The emission of
especially the low excitation lines is now boosted by a thin layer of
warm H$_2$O in the upper part of the disk with lower densities,
leading to a much larger discrepancy between the non-LTE and LTE case
due to subthermal excitation. To boost the line-to-continuum ratio
even more, the gas-to-dust ratio in Models 4 and 5 was increased by a
factor of 10 with respect to Model 1 through 3. The low excitation
lines in Model 4 now become too bright and therefore a
speculative vertical cold finger effect was added in Model 5. In this case, the
difference between the LTE and non-LTE spectrum is minimized because
the regions of the disk that have the largest deviation from LTE
contain little/no water vapor. In summary, 
beyond a radius $R\sim 0.3$~AU, levels having transitions in the
{\it Spitzer} IRS wavelength range are sub-thermally excited, leading
to biased estimates of the temperature and density distribution,
emitting areas and interpretation of high resolution ($\sim 3$~km/s)
line profiles when using LTE models.

2) In order to obtain more line emission from high excitation water
lines ($E_{up}>3000$~K), we introduced a steep gas temperature
increase at high altitudes in the disk, which is motivated by both
observations and models of, e.g., the [NeII] 12.81~$\mu$m and [OI]
6300~\AA\ lines. This increases the line strenghts by 20-100 percent,
and the low excitation lines are boosted most. This is not enough,
however, to reproduce the observed {\it Spitzer} IRS line-to-continuum
ratios.

3) The gas-to-dust ratio must be increased by one to two orders of
magnitude with respect to the canonical ISM ratio of $\sim 100-200$ in
order to approach the observed line strengths and line-to-continuum
ratios. An increase in gas-to-dust ratio lowers the dust opacity,
allowing lines to be formed at higher densities deeper in the disk.
An increase in the apparent gas-to-dust ratio can be explained by
grain growth and dust settling. Increasing the gas-to-dust ratio is a
more efficient way to increase line-to-continuum ratios than the
decoupling of gas from dust at high altitudes in the disk because the
former modification emphasizes high density gas.

4) Chemical models \citep[e.g.,][]{Glassgold2009, Woitke2009} show
that it is possible to obtain high water abundances in the transition
zone from hot ($T\sim 5000$~K) to cold gas ($T\sim 100-200$~K) and
that the range along the vertical axis where the maximum abundance
occurs becomes smaller with radius. Below this vertical zone, closer
to the midplane, water is less efficiently formed because the
neutral-neutral reactions that form it (O + H$_2$ $\rightarrow$ OH + H
followed by OH + H$_2$ $\rightarrow$ H$_2$O + H) contain activation
barriers. Even so, the predicted lower limit to the water abundance
($x_{\rm H_2O}\sim 10^{-6}$), or a suppression by 1-2 orders of
magnitude, still produces too much emission in the optically thick low
excitation lines.  Rather, a sharp drop in the surface water abundance
by up to 4 orders of magnitude beyond $\sim 1~$AU better reproduces
the observations. This sharp drop is difficult to explain with only
static chemical models, although it is not completely ruled out that
additional chemical porcesses could effectively reduce the water
abundance. We speculate a vertical cold finger effect, a process in
which water vapor is diffusively transported in the vertical direction
to a point below the snow line where the water freezes out and takes
part in the general growth and settling of solids to the midplane, may
provide a viable explanation (Fig. \ref{vert_coldfinger}). If this
effect does occur and is efficient, the radial distribution of water
line emission from the disk surface can become a direct tracer of the
location of the midplane snow line.

A chemical suppression of the water abundance, as compared a vertical
cold finger process would give rise to two completely different disk
surface chemistries as a function of radius that can be tested
observationally: If vertical transport is not important and chemical
processes along determine the surface water abundance, the chemistry
is oxygen dominated. Conversely, a vertical cold finger effect will
dramatically reduce the total oxygen abundance in the surface, giving
rise a carbon-dominated chemistry resulting in very different
radial distributions of organics and other species as observed in
mid-infrared surface molecular tracers. To test this, sensitive high
resolution mid-infrared spectroscopy is required, such as will be
offered by the James Webb Space Telescope (JWST) and the next
generation of Extremely Large Telescopes (ELTs), combined with
detailed excitation modeling.
  
The present study has qualitatively determined the water temperature and
abundance distribution required to explain observed mid-infrared
spectra of T Tauri disks. However, several important aspects that may
contribute to the formation and appearance of mid-infrared water lines
from protoplanetary disks are not discussed in this paper.

(i) Intrinsic line broadening: Here we adopted an intrinsic linewidth
that is dominated by thermal broadening. It is possible that
additional broadening from MRI driven turbulence would be able to
double the intrinsic linewidth, although the small predicted values
for turbulent viscosities would indicate that intrinsic line widths
are indeed dominated by thermal broadening. An increase of the
intrinsic line broadening would reduce the opacities in the lines,
thus increasing the effective critical densities, leading to lowered
intensities of the high excitation lines.

(ii) Stellar properties: The central stars in the observed {\it
  Spitzer} sample have masses ranging $M = 0.3$ to $3.0$~M$_\odot$,
and ages ranging from 1 to 10 Myrs, corresponding to luminosities and
effective temperatures ranging from $L=0.1$ to $20$~L$_\odot$ and
$T_{\rm eff}=2000-10000$~K, respectively. The modeling presented here
has not explored dependencies on the properties of the central star.

(iii) Inner holes: The maximum dust temperature will decrease when the
inner rim is located at larger radii, meaning that the high excitation
lines will become less prominent in disks with inner holes, while some
lower excitation lines may actually become brighter (see Pontoppidan
et al. 2009).

(iv) The model setup can easily be extended to other molecular species.
Work is underway to treat other observed molecular tracers in the
mid-infrared, including CO, HCN, C$_2$H$_2$, OH, etc.

These aspects will be discussed in a subsequent paper dedicated to a
full parameter study (Meijerink et al. 2009, in prep.).

\acknowledgments{ RM has been supported by NSF grant AST-0708922 to
  Caltech. Support for KMP was provided by NASA through Hubble
  Fellowship grant \#01201.01 awarded by the Space Telescope Science
  Institute, which is operated by the Association of Universities for
  Research in Astronomy, Inc., for NASA, under contract NAS
  5-26555. DRP is supported by the Marie Curie research Training
  Network 'Constellation' under grant no. MRTN-CT-2006-035890. The
  authors thank M. Spaans, A.G.G.M. Tielens, A. Glassgold, P. Woitke
  and I. Kamp for valuable discussions on the topic. }

\bibliographystyle{apj} 

\begin{thebibliography}{39}
\expandafter\ifx\csname natexlab\endcsname\relax\def\natexlab#1{#1}\fi

\bibitem[{{Ag{\'u}ndez} {et~al.}(2008){Ag{\'u}ndez}, {Cernicharo}, \&
  {Goicoechea}}]{Agundez2008}
{Ag{\'u}ndez}, M., {Cernicharo}, J., \& {Goicoechea}, J.~R. 2008, \aap, 483,
  831

\bibitem[{{Brown}(2007)}]{Brown2007b}
{Brown}, J. 2007, PhD thesis, California Institute of Technology

\bibitem[{{Carr} \& {Najita}(2008)}]{Carr2008}
{Carr}, J.~S., \& {Najita}, J.~R. 2008, Science, 319, 1504

\bibitem[{{Chiang} \& {Goldreich}(1997)}]{Chiang1997}
{Chiang}, E.~I., \& {Goldreich}, P. 1997, \apj, 490, 368

\bibitem[{{Ciesla}(2009)}]{Ciesla2009}
{Ciesla}, F.~J. 2009, Icarus, 200, 655

\bibitem[{{Ciesla} \& {Cuzzi}(2006)}]{Ciesla2006}
{Ciesla}, F.~J., \& {Cuzzi}, J.~N. 2006, Icarus, 181, 178

\bibitem[{{Dubernet} \& {Grosjean}(2002)}]{Dubernet2002}
{Dubernet}, M.-L., \& {Grosjean}, A. 2002, \aap, 390, 793

\bibitem[{{Dullemond} \& {Dominik}(2004)}]{Dullemond2004}
{Dullemond}, C.~P., \& {Dominik}, C. 2004, \aap, 417, 159

\bibitem[{{Ercolano} {et~al.}(2008){Ercolano}, {Drake}, {Raymond}, \&
  {Clarke}}]{Ercolano2008}
{Ercolano}, B., {Drake}, J.~J., {Raymond}, J.~C., \& {Clarke}, C.~C. 2008,
  \apj, 688, 398

\bibitem[{{Faure} {et~al.}(2007){Faure}, {Crimier}, {Ceccarelli}, {Valiron},
  {Wiesenfeld}, \& {Dubernet}}]{Faure2007}
{Faure}, A., {Crimier}, N., {Ceccarelli}, C., {Valiron}, P., {Wiesenfeld}, L.,
  \& {Dubernet}, M.~L. 2007, \aap, 472, 1029

\bibitem[{{Faure} {et~al.}(2004){Faure}, {Gorfinkiel}, \&
  {Tennyson}}]{Faure2004}
{Faure}, A., {Gorfinkiel}, J.~D., \& {Tennyson}, J. 2004, \mnras, 347, 323

\bibitem[{{Faure} \& {Josselin}(2008)}]{Faure2008}
{Faure}, A., \& {Josselin}, E. 2008, \aap, 492, 257

\bibitem[{{Fraser} {et~al.}(2001){Fraser}, {Collings}, {McCoustra}, \&
  {Williams}}]{Fraser2001}
{Fraser}, H.~J., {Collings}, M.~P., {McCoustra}, M.~R.~S., \& {Williams}, D.~A.
  2001, \mnras, 327, 1165

\bibitem[{{Glassgold} {et~al.}(2009){Glassgold}, {Meijerink}, \&
  {Najita}}]{Glassgold2009}
{Glassgold}, A.~E., {Meijerink}, R., \& {Najita}, J.~R. 2009, \apj, 701, 142

\bibitem[{{Glassgold} {et~al.}(1997){Glassgold}, {Najita}, \&
  {Igea}}]{Glassgold1997}
{Glassgold}, A.~E., {Najita}, J., \& {Igea}, J. 1997, \apj, 480, 344

\bibitem[{{Glassgold} {et~al.}(2004){Glassgold}, {Najita}, \&
  {Igea}}]{Glassgold2004}
---. 2004, \apj, 615, 972

\bibitem[{{Gorti} \& {Hollenbach}(2008)}]{Gorti2008}
{Gorti}, U., \& {Hollenbach}, D. 2008, \apj, 683, 287

\bibitem[{{Green} {et~al.}(1993){Green}, {Maluendes}, \& {McLean}}]{Green1993}
{Green}, S., {Maluendes}, S., \& {McLean}, A.~D. 1993, \apjs, 85, 181

\bibitem[{{Grosjean} {et~al.}(2003){Grosjean}, {Dubernet}, \&
  {Ceccarelli}}]{Grosjean2003}
{Grosjean}, A., {Dubernet}, M.-L., \& {Ceccarelli}, C. 2003, \aap, 408, 1197

\bibitem[{{Kamp} \& {Dullemond}(2004)}]{Kamp2004}
{Kamp}, I., \& {Dullemond}, C.~P. 2004, \apj, 615, 991

\bibitem[{{Kessler-Silacci} {et~al.}(2006){Kessler-Silacci}, {Augereau},
  {Dullemond}, {Geers}, {Lahuis}, {Evans}, {van Dishoeck}, {Blake}, {Boogert},
  {Brown}, {J{\o}rgensen}, {Knez}, \& {Pontoppidan}}]{Kessler2006}
{Kessler-Silacci}, J., {Augereau}, J.-C., {Dullemond}, C.~P., {Geers}, V.,
  {Lahuis}, F., {Evans}, II, N.~J., {van Dishoeck}, E.~F., {Blake}, G.~A.,
  {Boogert}, A.~C.~A., {Brown}, J., {J{\o}rgensen}, J.~K., {Knez}, C., \&
  {Pontoppidan}, K.~M. 2006, \apj, 639, 275

\bibitem[{{Kurucz}(1993)}]{Kurucz1993}
{Kurucz}, R.~L. 1993, in Astronomical Society of the Pacific Conference Series,
  Vol.~44, IAU Colloq. 138: Peculiar versus Normal Phenomena in A-type and
  Related Stars, ed. M.~M. {Dworetsky}, F.~{Castelli}, \& R.~{Faraggiana},
  87--+

\bibitem[{{Markwick} {et~al.}(2002){Markwick}, {Ilgner}, {Millar}, \&
  {Henning}}]{Markwick2002}
{Markwick}, A.~J., {Ilgner}, M., {Millar}, T.~J., \& {Henning}, T. 2002, \aap,
  385, 632

\bibitem[{{Meijerink} {et~al.}(2008){Meijerink}, {Poelman}, {Spaans},
  {Tielens}, \& {Glassgold}}]{Meijerink2008}
{Meijerink}, R., {Poelman}, D.~R., {Spaans}, M., {Tielens}, A.~G.~G.~M., \&
  {Glassgold}, A.~E. 2008, \apjl, 689, L57

\bibitem[{{Morrison} \& {McCammon}(1983)}]{Morrison1983}
{Morrison}, R., \& {McCammon}, D. 1983, \apj, 270, 119

\bibitem[{{Nomura} {et~al.}(2007){Nomura}, {Aikawa}, {Tsujimoto}, {Nakagawa},
  \& {Millar}}]{Nomura2007}
{Nomura}, H., {Aikawa}, Y., {Tsujimoto}, M., {Nakagawa}, Y., \& {Millar}, T.~J.
  2007, \apj, 661, 334

\bibitem[{{Nomura} \& {Millar}(2005)}]{Nomura2005}
{Nomura}, H., \& {Millar}, T.~J. 2005, \aap, 438, 923

\bibitem[{{Phillips} {et~al.}(1996){Phillips}, {Maluendes}, \&
  {Green}}]{Phillips1996}
{Phillips}, T.~R., {Maluendes}, S., \& {Green}, S. 1996, \apjs, 107, 467

\bibitem[{{Poelman} \& {Spaans}(2005)}]{Poelman2005}
{Poelman}, D.~R., \& {Spaans}, M. 2005, \aap, 440, 559

\bibitem[{{Poelman} \& {Spaans}(2006)}]{Poelman2006}
---. 2006, \aap, 453, 615

\bibitem[{{Pontoppidan}(2006)}]{Pontoppidan2006}
{Pontoppidan}, K.~M. 2006, \aap, 453, L47

\bibitem[{{Pontoppidan} {et~al.}(2009){Pontoppidan}, {Meijerink}, {Dullemond},
  \& {Blake}}]{Pontoppidan2009}
{Pontoppidan}, K.~M., {Meijerink}, R., {Dullemond}, C.~P., \& {Blake}, G.~A.
  2009, ArXiv:0908.2997

\bibitem[{{Raymond} {et~al.}(2004){Raymond}, {Quinn}, \&
  {Lunine}}]{Raymond2004}
{Raymond}, S.~N., {Quinn}, T., \& {Lunine}, J.~I. 2004, Icarus, 168, 1

\bibitem[{{Rothman} {et~al.}(2005){Rothman}, {Jacquemart}, {Barbe}, {Chris
  Benner}, {Birk}, {Brown}, {Carleer}, {Chackerian}, {Chance}, {Coudert},
  {Dana}, {Devi}, {Flaud}, {Gamache}, {Goldman}, {Hartmann}, {Jucks}, {Maki},
  {Mandin}, {Massie}, {Orphal}, {Perrin}, {Rinsland}, {Smith}, {Tennyson},
  {Tolchenov}, {Toth}, {Vander Auwera}, {Varanasi}, \& {Wagner}}]{Rothman2005}
{Rothman}, L.~S., {Jacquemart}, D., {Barbe}, A., {Chris Benner}, D., {Birk},
  M., {Brown}, L.~R., {Carleer}, M.~R., {Chackerian}, C., {Chance}, K.,
  {Coudert}, L.~H., {Dana}, V., {Devi}, V.~M., {Flaud}, J.-M., {Gamache},
  R.~R., {Goldman}, A., {Hartmann}, J.-M., {Jucks}, K.~W., {Maki}, A.~G.,
  {Mandin}, J.-Y., {Massie}, S.~T., {Orphal}, J., {Perrin}, A., {Rinsland},
  C.~P., {Smith}, M.~A.~H., {Tennyson}, J., {Tolchenov}, R.~N., {Toth}, R.~A.,
  {Vander Auwera}, J., {Varanasi}, P., \& {Wagner}, G. 2005, Journal of
  Quantitative Spectroscopy and Radiative Transfer, 96, 139

\bibitem[{{Salyk} {et~al.}(2008){Salyk}, {Pontoppidan}, {Blake}, {Lahuis}, {van
  Dishoeck}, \& {Evans}}]{Salyk2008}
{Salyk}, C., {Pontoppidan}, K.~M., {Blake}, G.~A., {Lahuis}, F., {van
  Dishoeck}, E.~F., \& {Evans}, II, N.~J. 2008, \apjl, 676, L49

\bibitem[{{Siess} {et~al.}(2000){Siess}, {Dufour}, \& {Forestini}}]{Siess2000}
{Siess}, L., {Dufour}, E., \& {Forestini}, M. 2000, \aap, 358, 593

\bibitem[{{Stevenson} \& {Lunine}(1988)}]{Stevenson1988}
{Stevenson}, D.~J., \& {Lunine}, J.~I. 1988, Icarus, 75, 146

\bibitem[{{Woitke} {et~al.}(2009){Woitke}, {Kamp}, \& {Thi}}]{Woitke2009}
{Woitke}, P., {Kamp}, I., \& {Thi}, W.-F. 2009, \aap, 501, 383

\bibitem[{{Woods} \& {Willacy}(2009)}]{Woods2009}
{Woods}, P.~M., \& {Willacy}, K. 2009, \apj, 693, 1360

\end{thebibliography}

\end{document}